\documentclass[11pt]{article}

\usepackage{xcolor}
\usepackage{ulem}
\usepackage{eps fig}
\usepackage{amsfonts}
\usepackage{latexsym}
\usepackage{amsmath,amssymb,bm}
\usepackage{verbatim}
\usepackage{color}
\usepackage{hyperref}
\usepackage{cite}

\usepackage{setspace}
\usepackage[textheight=9in, textwidth=6.5in, letterpaper]{geometry}
\def\half{{1\over 2}}
\numberwithin{equation}{section}

 \def\p{\partial}

\newcommand{\bea}{\begin{eqnarray}}
\newcommand{\eea}{\end{eqnarray}}
\newcommand{\be}{\begin{equation}}
\newcommand{\ee}{\end{equation}}
\newcommand{\ba}{\begin{align}}
\newcommand{\ea}{\end{align}}

\newcommand{\Pol}{\text{Pol}(\Delta)}

  \makeatletter
  \let\over=\@@over \let\overwithdelims=\@@overwithdelims
  \let\atop=\@@atop \let\atopwithdelims=\@@atopwithdelims
  \let\above=\@@above \let\abovewithdelims=\@@abovewithdelims

\renewcommand\section{\@startsection {section}{1}{\z@}%
                                   {-3.5ex \@plus -1ex \@minus -.2ex}
                                   {2.3ex \@plus.2ex}%
                                   {\normalfont\large\bfseries}}

\renewcommand\subsection{\@startsection{subsection}{2}{\z@}%
                                     {-3.25ex\@plus -1ex \@minus -.2ex}%
                                     {1.5ex \@plus .2ex}%
                                     {\normalfont\bfseries}}

\def\half{{1 \over 2}}

\def\Or[#1]{{\text{O}}\left({#1}\right)}
\def\dotl[#1,#2]{\left\langle #1, #2 \right\rangle}
\def\dotlb[#1,#2]{[ #1, #2 ]}
\def\dotp[#1,#2]{(#1) \cdot (#2)}
\def\aff[#1,#2]{\hat{#1}(#2)}
\def\n4sym{{\cal N}=4 SYM}
\def\>{\rangle}
\def\<{\langle}

\def\weight[#1,#2,#3]{\{(#1),#2,#3\}}
\def\ads[#1]{$\text{AdS}_{#1}$}

\newcommand{\nn}{\nonumber}
\newcommand{\ZZZ}{\mathbb{Z}}
\newcommand{\cO}{{\cal{O}}}
\newcommand{\beq}{\begin{equation}}
\newcommand{\eeq}{\end{equation}}

\begin{document}

\unitlength = 1mm

\pagestyle{empty}

\setcounter{page}{0}
\pagestyle{plain}

\setcounter{tocdepth}{3}

\def\vx{{\vec x}}
\def\ip{${\cal I}^+$}
\def\p{\partial}
\def\po{$\cal P_O$}

\thispagestyle{empty}
\begin{center}
{  \Large{\textsc{Partition Functions in Even Dimensional AdS \\via Quasinormal Mode Methods}} }
\vspace*{0.25cm}

Cynthia Keeler$^{a}$
and Gim Seng Ng$^{b}$
\vspace*{0.25cm}

{\it $^a$Department of Physics, University of Michigan,
Ann Arbor, MI-48109, USA}

{\it $^b$Center for the Fundamental Laws of Nature, Harvard University, Cambridge, MA 02138, USA}
\end{center}

\begin{abstract}
In this note, we calculate the one-loop determinant for a massive
scalar (with conformal dimension $\Delta$) in even-dimensional AdS$_{d+1}$ space, using the quasinormal
mode method developed in \cite{Denef:2009kn} by Denef, Hartnoll, and Sachdev.
Working first in two dimensions on the related Euclidean hyperbolic
plane $H_2$, we find a series of zero modes for negative real values of
$\Delta$ whose presence indicates a series of poles in the one-loop
partition function $Z(\Delta)$ in the $\Delta$ complex plane; these
poles contribute temperature-independent terms to the thermal AdS
partition function computed in \cite{Denef:2009kn}.  Our results match those in a series of papers by
Camporesi and Higuchi, as well as Gopakumar et.al.\cite{Gopakumar:2011qs} and Banerjee et.al.\cite{Banerjee:2010qc}.
We additionally examine the meaning of these zero modes, finding that
they Wick-rotate to quasinormal modes of the AdS$_2$ black hole.  They
are also interpretable as matrix elements of the discrete series
representations of $SO(2,1)$ in the space of smooth functions on
$S^1$.
We generalize our results to general even dimensional AdS$_{2n}$, again
finding a series of zero modes which are related to discrete series
representations of $SO(2n,1)$, the motion group of $H_{2n}$.
\end{abstract}
\newpage
\tableofcontents

\section{Introduction}

Quantum effects in AdS spacetimes are of great interest to the theoretical physics community, particularly since the advent of the AdS/CFT correspondence nearly two decades ago.  Even leading one-loop effects can contain important physical results. For example, quantum corrections to the entropy of black holes can be computed using the quantum entropy function \cite{Sen:2008vm,Banerjee:2010qc,Mandal:2010cj}, which requires the one-loop effective action due to the matter fields on an AdS$_2 \times M$ spacetime.

As such, there is interest in developing efficient methods for various one-loop calculations.  Several methods for calculating the one-loop determinant for a particular field fluctuation are already known, mostly based on the heat kernel method (see e.g. \cite{Vassilevich:2003xt}).
There are several known ways to calculate the heat kernel.  First, if one is only interested in large mass  behavior (for a massive field of mass $m$), the curvature expansion in terms of local invariants is known for many field types up to 10th order \cite{Vassilevich:2003xt}.
If we wish to know the heat kernel outside of a $1/m$ expansion, then we can expand the heat kernel as a sum over normalized eigenfunctions of the Laplacian weighted by their eigenvalues; this is the approach used in the series of papers \cite{Camporesi:1991nw,Camporesi:1992wn,Camporesi199457,Camporesi:1994ga} and again by \cite{Banerjee:2010qc} for AdS$_2 \times S^2$ spacetimes.
These eigenfunctions are generally special functions and the details, particularly for higher spin fields, can be challenging.

Of course these eigenfunctions of the Laplacian themselves arise from the group theory structure of the spacetime; the Laplacian is the quadratic Casimir for the motion group of the space.
As such Gopakumar et. al. developed a method for computing the heat kernel from group theory characters \cite{David:2009xg}, which they extended to AdS$_{d+1}$ for $d\geq 2$ in \cite{Gopakumar:2011qs}. This powerful method works well for spacetimes where we understand the group theory structure.

In two papers \cite{Denef:2009kn,Denef:2009yy}, Denef, Hartnoll and Sachdev developed a method\footnote{
Although we will primarily discuss the case of AdS$_{2n}$ in this note, this method actually applies to the case of de Sitter space as well, as was discussed in the original paper of  \cite{Denef:2009kn}. It would be interesting to understand
the relations between the interpretation of de Sitter quasinormal modes as highest-weight representations\cite{Ng:2012xp,Jafferis:2013qia} and their roles in this 1-loop partition function method.
}
 not directly based on the heat kernel approach.
  In this method, developed from an argument in \cite{Coleman1988} (also see \cite{Hartman:2006dy,Aros:2009pg}), the partition function $Z({\Delta})$ (of a massive scalar of mass $m$) is calculated via a complex analysis trick\footnote{This method has been generalized to fields of arbitrary spin on AdS$_3$ in \cite{Datta:2011za,Datta:2012gc,Zhang:2012kya}.}.  Rather than specifying a physical mass $m$ (corresponding to a particular conformal dimensional $\Delta$), and then calculating the partition function at that mass, this method studies $Z(\Delta)$ as a function in the complex $\Delta$ plane.  If we assume $Z(\Delta)$ is a meromorphic function in complex $\Delta$ plane%
\footnote{This assumption is reasonable; we will apply boundary conditions in terms of $\Delta$, and there is no reason to expect a branch cut or essential singularity in $Z(\Delta)$.}%
, then to find the function completely we need only the location and multiplicities of its poles and zeros, as well as one (entire) function which we term $\exp[ \Pol]$, where $\Pol$ is some polynomial of $\Delta$.  In the case of $Z(\Delta)$, this extra function is completely constrained by the behavior of the function at large mass; that is, we can use the heat kernel curvature expansion to completely determine $Z(\Delta)$.

As shown in \cite{Denef:2009kn}, for scalars $Z(\Delta)$ has only poles; these poles are due to zero modes in Euclidean space, which Wick-rotate to quasinormal modes in Lorentzian signature (or normal modes in global Lorentzian AdS). As such we will use the term ``zero mode method'' when working in Euclidean signature, but ``quasinormal mode method'' when working in Lorentzian signature. In fact \cite{Denef:2009kn} does provide a formula for thermal AdS using a ``normal mode'' method, referencing the global normal modes. For odd-dimensional AdS spacetimes, these modes are sufficient and, as shown for AdS$_3$ in \cite{Denef:2009kn}, they correctly reproduce the partition function in \cite{Giombi:2008vd}.

However, as we will show in this paper, on even-dimensional AdS spacetimes the normal modes are insufficient.
Although the normal modes capture all {\it temperature-dependent} information (which is often the behavior of primary interest), if one wants to obtain the {\it temperature-independent} piece, the normal modes are insufficient.  Since the temperature-independent piece contains logarithmic behavior, its presence is needed to show $\Pol$ is truly polynomial. To properly compute this piece, it is necessary to include a set of modes whose boundary ``falloff'' is set by $\Delta$. However these modes are nonetheless not normalizable, because they are located in the complex $\Delta$ plane on the negative real axis.  In the case of AdS$_2$, we will see that they are still interpretable as a set of quasinormal modes.

In Section \ref{d+1=2section}, we apply the zero mode method to compute the scalar one-loop determinant on Euclidean AdS$_2$, or $H_2$. In the process we review the details of the method.  We then compare our results to previous results, finding agreement.  In Section \ref{funthings}, we explore the physical and mathematical meaning of the novel zero modes we have found, with the hope of better understanding when such modes might be present in other spacetimes. In Section \ref{thermalsection}, we study the thermal AdS$_2$ as a quotient space of $H_2$.  We show our results from Section \ref{d+1=2section} can be combined with those of \cite{Denef:2009kn} to match the thermal partition function as computed via the methods of \cite{Gopakumar:2011qs}, including the temperature independent piece.
In Section \ref{2nsection}, we extend our analysis to general even-dimensional AdS spaces, again matching with previous results.  In the conclusion, we discuss application of this method to odd-dimensional AdS spaces, and also highlight important caveats to consider when applying the zero/quasinormal mode method.  The appendix contains further calculational details.


\section{Applying the Zero-mode Method to AdS$_2$}\label{d+1=2section}
In this section, we use the zero-mode method to calculate the $1$-loop partition function for a complex scalar $\phi$ with mass $m$ in Euclidean AdS$_2$ with AdS length $L$.  We begin by setting notation and briefly reviewing the zero-mode method from \cite{Denef:2009kn}. We will extend their method for compact spaces to the non-compact AdS$_2$.

We work in coordinates
\be\label{AdS2EucDisc}
ds_E^2= L^2 \left(d\eta^2 +\sinh^2 \eta \, d\theta^2\right), \qquad \theta \sim \theta+ 2 \pi,\quad
 \eta\ge 0,
\ee  where $\eta\rightarrow \infty$ is the boundary.
Writing the partition function in terms of the scalar determinant we have:
\be
\label{ZDeltadef}
Z(\Delta) = \int {\cal D}\phi \, e^{- \int \phi^*(-\nabla^2+m^2)\phi} \propto \frac{1}{\det \left(-\nabla^2+m^2\right)},
\ee
where $\Delta$ solves $(mL)^2 = \Delta(\Delta-1)$. Here, we choose the root $\Delta$ according to the standard Dirichlet boundary condition for AdS$_2$, i.e. we impose the Dirichlet-type boundary condition
\begin{align} \label{phibc}
\phi &\sim \left(\sinh\eta\right)^{-\Delta} \text{ for } \eta \rightarrow \infty,
\nn\\
\Delta &\equiv \frac{1}{2}+\sqrt{\frac{1}{4}+m^2 L^2} \equiv \frac{1}{2}+\mu,
\end{align}
at the AdS boundary.

In order to calculate the determinant, we will use the method developed in \cite{Denef:2009kn}.
If $Z$ is a meromorphic function of $\Delta$, then we can use the Weierstrass factorization theorem%
\footnote{
The initial version of the Weierstrass factorization theorem only applies to analytic functions, but it can be extended to meromorphic functions.  Additionally convergence of infinite products will not concern us, so we do not worry about using the proper Weierstrass factors.  If we include information from the growth behavior at large $\Delta$, which we will have from the heat kernel curvature expansion, we could use instead the Hadamard factorization theorem; here we will not need these complications.
}
 from complex analysis to find the function.  Under this theorem, if we know the poles and zeros (and their multiplicities) of a meromorphic function, then we know the function itself, up to a factor which is an entire function.  The determinant we are interested in will have no zeros, only poles, so we can write
\be\label{Zaspoles}
Z(\Delta) = e^{\Pol} \prod_i \left(\Delta-\Delta_i\right)^{-d_i},
\ee
where the $\Delta_i$ are poles (of $Z(\Delta)$) each with degeneracy $d_i$. Importantly, $\Pol$ is a polynomial in $\Delta$, since $e^{\Pol}$ is entire.

We find the poles in $Z(\Delta)$ by searching for values of $\Delta$ where $\det \left(-\nabla^2 + m^2\right)$ becomes zero. These $\Delta_\star$ occur whenever there exists a $\phi_\star$ solving
\be\label{zeromodeeq}
\left(-\nabla^2 + \frac{\Delta_\star(\Delta_\star-1)}{L^2}\right)\phi_\star=0,
\ee
where $\phi_\star$ is smooth in the interior of the (Euclidean) AdS, and satisfies the boundary condition (\ref{phibc}).  That is, at these (generally complex) $\Delta_\star$, $\phi_\star$ is a zero mode.  Thus at $\Delta=\Delta_\star$, $Z(\Delta)$ has a pole; the degeneracy of this pole is the number of smooth solutions to (\ref{zeromodeeq}) which also satisfy (\ref{phibc}). For a spacetime with a compact direction, these $\Delta_\star$ can be labelled by a {\it discrete} index $i$; we call this set the $\Delta_i$.  As we will show for AdS$_2$, even a noncompact space may have a discrete set of poles, when we choose the appropriate boundary condition.

As shown in \cite{Denef:2009kn}, for spacetimes which are Euclidean black holes, these zero modes Wick rotate to become quasinormal modes of the corresponding Lorentzian black holes.  We will discuss the details of this Wick rotation in the Euclidean AdS spacetime in Section \ref{Wickrotation} below.

As a preview, the zero-mode method for calculating the $1$-loop determinant has three steps, which we will explain in detail as we proceed:
\begin{itemize}
\item Find $\Delta_i$ where there exists a zero mode $\phi_{i}$, and the multiplicities $d_i$ of these zero modes.
\item Use Eq.~(\ref{Zaspoles}) to express $Z(\Delta)$ in terms of these $\Delta_i$ and $d_i$.  Zeta-function regularize $\log Z(\Delta)$ to perform the infinite sum.
\item Find the polynomial $\Pol$ by matching the zeta-function regularization at large $\Delta$ to the local heat kernel curvature expansion expression for $\log Z(\Delta)$.
\end{itemize}

\subsection{Finding the zero modes}\label{zeromodesection}
The first step is to find the zero modes in the complex $\Delta$ plane. Following \cite{Camporesi:1994ga}, the solutions to (\ref{zeromodeeq}) which are regular at $\eta =0$ in Euclidean AdS$_2$ can be written as
\be \label{AdS2modes}
\phi_{\Delta l}=  e^{il\theta}(i \sinh \eta)^{|l|} F\left[\Delta+|l|, |l|+1-\Delta;|l|+1;-\sinh^2 \left(\frac{\eta}{2}\right)\right],
\ee
where $F$ is the hypergeometric function also written $\hspace{-.15cm}\phantom{.}_{2}F_1$.  We have already imposed part of the boundary condition at the origin by choosing the solution of the hypergeometric equation which is regular at $\eta =0$.  To impose that $\phi_{\Delta l}$ be single-valued due to periodicity of $\theta\sim\theta+2\pi$, we additionally impose $l \in \ZZZ$.

Now, rather than imposing delta-function normalizability as in \cite{Camporesi:1994ga}, we impose the boundary condition by insisting the modes have only the Dirichlet-like behavior $(\sinh \eta)^{-\Delta}$ as $\eta \rightarrow \infty$. That is, we wish to turn the behavior $(\sinh\eta)^{\Delta-1}$ off entirely%
\footnote{We use this boundary condition in accordance with AdS/CFT intuition.  We will show in Section \ref{fulleigen} below that this choice agrees with the results from the heat kernel eigenfunction expansion done over delta-function normalizable modes.  The agreement is a consistency check that our boundary condition produces the same result. It would also be interesting to understand why these two prescriptions are equivalent.}%
. Using the result (\ref{d+1=2cond}) from  Appendix \ref{proofofzeromode}, this is accomplished whenever $\Delta=\Delta_\star$ where
\be\label{AdS2zeros}
\Delta_\star \in \ZZZ_{\leq 0}, \qquad |l| \leq -\Delta_\star.
\ee
In other words, at each nonpositive integer $\Delta_\star$, we have a zero mode of multiplicity $-2\Delta_\star +1$. There are no other zero modes for any $\Delta$ in the complex plane.  Since these zero modes occur at integer values $\Delta_\star$, they are discrete. In order for the zero-mode method to work, we require a discrete set of poles; if the set of poles had an accumulation point (other than infinity), the Weierstrass theorem would not apply.

As we will discuss further in Section \ref{funthings} below, these modes have several interesting physical and mathematical properties.  They are the Wick-rotation of quasinormal modes for AdS$_2$ black holes, as well as the manifestation of the discrete series representation of the $SO(2,1)$ symmetry of AdS$_2$, on the space of smooth functions on the circle $S^1$.

\subsection{Summing the contributions of the zero modes}
For now, as we have found a discrete set of zero modes, we continue our calculation of the AdS$_2$ complex scalar partition function by evaluating $\log$ of (\ref{Zaspoles})%
\footnote{
One might recognize these expressions as similar to that of the computations of dS$_2$ in \cite{Denef:2009kn}.  There are two important differences: First, due to the AdS Dirichlet boundary condition, we are only summing over the $\Delta$ boundary condition here (and not $1-\Delta$). Second, the relationship between $\Delta$ and the mass  for AdS differs by a sign from that for de Sitter.
}:
\begin{align}
\log Z(\Delta) &= \Pol +\sum_{\Delta_\star\in \ZZZ_{\leq 0}}(2\Delta_\star-1)\log\left(\Delta-\Delta_\star\right),
\nn\\
 &= \Pol-\sum_{k=0}^{\infty}(2k+1)\log(\Delta+k),
\nn \\\label{Zinzetas}
 &= \Pol+2\zeta'(-1,\Delta)-(2\Delta-1)\zeta'(0,\Delta).
\end{align}
where we have used zeta function regularization as in \cite{Denef:2009kn}%
\footnote{More complicated spacetimes might require use of a generalized zeta function, such as the multiple Barnes zeta function, but we will find the Hurwitz function sufficient here.}%
. Here $\zeta(s,x)$ is the Hurwitz zeta function defined by analytic continuation of $\zeta(s,x)= \sum_{k=0}^{\infty} (x+k)^{-s}$, and $\zeta'(s,x)\equiv \partial_s \zeta(s,x)$, whose large $\Delta$ behavior is given in \cite{Denef:2009kn} (which is taken from Appendix A of \cite{2001NuPhB.612..492D})
as:
\be\label{Zetaexpand}
\zeta(s,\Delta) \simeq \frac{1}{\Gamma(s)}\sum_{k=0}^{\infty} \frac{(-1)^k B_k}{k!}\frac{\Gamma(k+s-1)}{\Delta^{k+s-1}},
\ee where $B_k$'s are the Bernoulli numbers.

Since we will find $\Pol$ by expanding the expression (\ref{Zinzetas}) at large $\Delta$ and comparing to $Z(\Delta)$ as found from the heat kernel curvature expansion, we now present the expansion of the zeta-function expression.  In order to compare more easily with the heat kernel expansion, we also expand in terms of $\mu=\Delta-1/2$ as in (\ref{phibc}). Expanding Eq.~(\ref{Zinzetas}) at large $\Delta$ in terms of $\mu$, using (\ref{Zetaexpand}), gives
\begin{align}\label{Zatlargemu}
\log Z (\Delta)&=\mbox{Pol}(\Delta)+\frac{3 }{2}\Delta ^2-\Delta
-\half\left(\Delta ^2-\Delta+\frac{1}{3}\right) \log (\Delta^2 )
 +\frac{1}{12 \Delta }
+\frac{1}{120 \Delta ^2}
-\frac{1}{360 \Delta ^3}
+\cO(\Delta^{-4}),
\nn\\
&=\mbox{Pol}(\mu)
+\frac{3}{2}\mu^2
-\frac{1}{2}\left(\mu^2+\frac{1}{12}\right)\log\mu^2-\frac{7}{940}\mu^{-2}+\cO(\mu^{-4}),
\end{align}
where we have replaced $\mbox{Pol}(\Delta)$ with $\mbox{Pol}(\mu)$, as a polynomial in $\Delta$ is also a polynomial in $\mu$.

\subsection{Finding $\Pol$ via local curvature invariants}
\label{findPolAdS2}
Now we still must compute $\Pol$, which we do by studying the behavior of $Z(\Delta)$ at large $\Delta$, via a heat kernel curvature expansion.  Specifically, we set $\Pol$ by requiring
\be\label{Zfromcurvature}
\log Z(\Delta) = \text{const.} + (4\pi)^{-\frac{d+1}{2}}\sum^{d+1}_{k=0} a_k \int_0^\infty \frac{dt}{t}t^{\frac{k-d-1}{2}}e^{-t m^2} + {\cal O}(m^{-1}),
\ee
where $a_k$ are the  kernel curvature coefficients, in our case for the operator $-\nabla^2$. When there is no boundary contribution,
 the curvature invariants $a_k=0$ for odd $k$ while for even $k$, $a_k\equiv\int d^{d+1}x \sqrt{g} a_k(x)$ can be expressed following \cite{Denef:2009kn} as
\begin{align}
\label{curvinvar}
a_0(x) =& 1,
\quad
a_2(x) =
\frac{R}{6},\quad
a_4(x) =
 \frac{1}{360}( 12 \nabla^2 R + 5 R^2
-2 R_{\mu\nu}R^{\mu\nu}+2R_{\mu\nu\rho\sigma}R^{\mu\nu\rho\sigma}),
\end{align}
where we have included $a_4$ for use in future sections.

Setting (\ref{Zfromcurvature}) equal to the large $\mu$ expansion (\ref{Zatlargemu}), we solve for $\Pol$, which must be a polynomial.  If it is not possible to set these two expressions equal by fixing $\Pol$ to be a polynomial, than either some zero modes have not been included in the sum, or the meromorphicity of $Z(\Delta)$ is suspect. On the other hand, if we can equate these expressions with a polynomial $\Pol$, then that provides a nontrivial check on our calculation.

Now we expand $\log Z(\Delta)$ at large mass using the heat kernel curvature expansion in terms of (\ref{Zfromcurvature}) and (\ref{curvinvar}).  Following \cite{Denef:2009kn}, we introduce a lower cutoff of $e^{-\gamma}\Lambda^{-2}$ on the heat kernel integral in (\ref{Zfromcurvature}), where $\gamma$ is the Euler number and $\Lambda$ labels the IR cut-off.
For the scalar operator we study,
\begin{align}
\log Z(\Delta)= &\frac{1}{4\pi}
\int_{H_2} d^2 x \sqrt{g}
\left(
 \int_{e^{-\gamma}\Lambda^{-2}}^\infty \frac{dt}{t^2} e^{-t m^2}
  + \frac{R}{6} \int_{e^{-\gamma}\Lambda^{-2}}^\infty \frac{dt}{t}e^{-t m^2}
  \right)
   + {\cal O}(m^{-1})
\nn\\
      = &\frac{1}{4\pi}
\int_{H_2} d^2 x \sqrt{g}
\left(
\frac{\mu^2-\frac{1}{4}}{L^2}
\log\frac{\mu^2}{e (L\Lambda_{RG})^2}
-\frac{1}{4L^2}
-\frac{R}{6}  \log \frac{\mu^2}{L^2 \Lambda_{RG}^2}
  \right)
\nn\\\label{likeDHS}
  &+\frac{1}{4\pi}
\int_{H_2} d^2 x \sqrt{g}
\left(
\frac{\mu^2-\frac{1}{4}}{L^2}
\log\frac{\Lambda^2_{RG}}{ \Lambda^2}
-\frac{R}{6}  \log \frac{\Lambda^2_{RG}}{\Lambda^2}
  \right)
   + {\cal O}(\mu^{-1})~ ,
\end{align}
 where we have dropped an $m$-independent quadratically divergent term and introduced an arbitrary renormalization scale $\Lambda_{RG}$.  As in \cite{Denef:2009kn}, we recognize the last line here as renormalization of the classical couplings.  We can thus account for their effect by using the logarithmically running couplings in evaluating the classical contribution to the action, and we do not need to consider these terms further here.

In order to compare the remaining expression with our zero-mode method result in Eq.~(\ref{Zatlargemu}), we need to evaluate the volume of the hyperbolic space. Following discussions in the context of holographic renormalization in AdS/CFT%
\footnote{
See, e.g., \cite{Diaz:2007an,Maldacena:2012xp,Giombi:2014iua}.}, %
the regularized Euclidean action involves the regularized volume of $H_{2n}$, which is given by
\be
\mbox{Vol}_{reg}(H_{2n})\equiv
\half (-L^2)^{n} V_{S^{2n}},\quad V_{S^{2n}}\equiv
\frac{2 \pi^{n+\half}}{\Gamma(n+\half)}.
\ee
With the regularized volume $\mbox{Vol}_{reg}(H_{2})=-2\pi L^2$, the first line in the second equality of Eq.~(\ref{likeDHS}) evaluates to be
\be\label{eq:PolfromHKH2}
\half\mu^2
-\half
\left(\mu^2+\frac{1}{12}\right)\log\mu^2
+\half\left(\mu^2+\frac{1}{12}\right)
\log( L\Lambda_{RG})^2,
\ee
where we have also used $R_{AdS_2}=-2/L^2$ . Comparing Eq.~(\ref{Zatlargemu}) and Eq.~(\ref{eq:PolfromHKH2}), we find $\Pol$ to be
\begin{align}
   \mbox{Pol}(\mu)
=&\left[-1+\log( L\Lambda_{RG})
\right]\mu^2+
\frac{1}{12}
\log( L\Lambda_{RG})
\nn\\
\mbox{or}\quad \mbox{Pol}(\Delta)
=& \left[-1+\log \left(L \Lambda _{RG}\right)\right] \Delta (\Delta-1)+\frac{1}{3} \log \left(L\Lambda _{RG}\right)-\frac{1}{4} ,
\end{align}
where in the second line we have reexpressed the polynomial in terms of $\Delta$.  Combining this result with (\ref{Zinzetas}), the final expression for $Z(\Delta)$ becomes
\be\label{Zfinalzero}
Z(\Delta) = -\Delta^2+\Delta -\frac{1}{4}+\left[\Delta^2-\Delta+\frac{1}{3}\right]\log\left(L \Lambda_{RG}\right)+2\zeta'(-1,\Delta)-(2\Delta-1)\zeta'(0,\Delta).
\ee
For latter comparison with the heat kernel method, we expand this final $Z$ in terms of $\mu$:
\be\label{Zfinalzero1}
Z(\mu)=
\frac{1}{2}\mu^2
-\frac{1}{2}\left(\mu^2+\frac{1}{12}\right)\log\mu^2
+\left(
\mu^2+\frac{1}{12}\right)\log( L\Lambda_{RG})
-\frac{7}{940}\mu^{-2}+
\cO(\mu^{-4}).
\ee

Before comparing with the full eigenfunction heat kernel computation, let us make few comments on the nature of our computation. Here we have actually computed the inverse of the one loop determinant, $1/\det \left(-\nabla^2+m^2\right)$, on Euclidean AdS$_2$ with a Dirichlet boundary condition.  This determinant is equivalent to the partition function for a complex scalar on the same space, via Gaussian integration, as long as we are evaluating the partition function away from a zero mode. If we need the value of $Z$ at one of the $\Delta_\star$, we would need to more carefully evaluate the path integral.

We also note that our calculation here is not just a direct extension of the prescription given in \cite{Denef:2009kn}, since the proper discrete set of modes to sum over in Euclidean AdS$_2$ is not a priori clear, as the space is non-compact. Here we have provided a prescription for finding this discrete set of modes.

\subsection{Comparison with Full Eigenfunction Heat Kernel Method}\label{fulleigen}

We now compare our result with previous results for the scalar partition function on AdS spacetimes.  In contrast to the direct computation of the partition function developed in \cite{Denef:2009kn} which we have adapted for Euclidean AdS$_{d+1}$, the other methods described here make use of the heat kernel on $H_{d+1}$.  More details about the heat kernel can be found in \cite{Vassilevich:2003xt}
while its application to $H_{d+1}$ can be found in \cite{Camporesi:1994ga}. The heat kernel satisfies
\be\label{Keqn}
(\partial_t-\square_x)K(x,x';t) = 0; \qquad K(x,x';0)=\delta^{(d+1)}(x-x'),
\ee
where $\square_x$ is the scalar Laplacian acting only on $x$, where $x$ is a point on AdS.  We can relate this object to the corresponding complex scalar partition function via
\begin{align}
\log Z(\Delta) \label{kappaline}
&= -\sum_n \log \kappa_n= \int_\epsilon^\infty \frac{dt}{t}\sum_n e^{-\kappa_n t-m^2 t}
\\\label{KtoZ}
 &= \int_\epsilon^\infty \frac{dt}{t} \int d^{d+1} x\, \sqrt{ g}\, K(x,x;t).
\end{align}
Here, $-\kappa_n$ are the eigenvalues of the scalar Laplacian, and the sum is done over these eigenvalues for a complete set of (delta-)normalizable modes.

Since the heat kernel satisfying equation (\ref{Keqn}) can be written in terms of an eigenvalue-weighted sum over a complete set of normalizable eigenstates, we can use this information to construct the heat kernel, and thus the partition function.  This is the method most closely followed by Camporesi and Higuchi in their series of papers \cite{Camporesi:1991nw,Camporesi:1992wn,Camporesi:1994ga,Camporesi199457}; it is also the method used in \cite{Banerjee:2010qc}.

For later usage, we will briefly generalize to Euclidean AdS$_{d+1}$, or $H_{d+1}$. In \cite{Camporesi:1994ga}, the eigenstates for $H_{d+1}$ which satisfy
\be
\square \phi = -\left(\lambda^2+\frac{d^2}{4}\right)\phi,
\ee
are written as
\be
 \phi=(i\sinh \eta)^l F\left[i\lambda+\frac{d}{2}+l, -i\lambda + \frac{d}{2}+l;l+\frac{d+1}{2};-\sinh^2 \frac{\eta}{2}\right] Y_{l \vec{m}},
\ee
where $F(a,b;c;z)$ is the hypergeometric function, the $Y_{l \vec{m}}$ are the orthonormal spherical harmonics on $S^d$ satisfying
\be
\nabla^2 Y_{l\vec{m}}= -l (l+d-1) Y_{l\vec{m}},
\ee
and $\vec{m}$ is a $(d-1)$-dimensional vector of integers distinguishing among distinct harmonics with the same eigenvalue.

 For real $\lambda$, these are the only eigenfunctions which are delta-function normalizable, and thus they can be used to compute the heat kernel via a sum over eigenstates. However, for real $\lambda$ these eigenfunctions do not obey the Dirichlet boundary condition at the boundary $H_{d+1}$, and thus here they are not candidates for the zero modes used in the zero-mode method of Section \ref{zeromodesection}.

 Since $\lambda$ is continuous, we must extend the sum over states in (\ref{kappaline}) to an integral; this is done via the spectral function $\tilde{\mu}(\lambda)$.  The coincident heat kernel then becomes
\be\label{eq:heatkerneldef}
 K(x,x;t)=\int^\infty_0 d\lambda \,\tilde{\mu}(\lambda) ~\exp{\left[-t\left(\lambda^2+\left(\frac{d}{2}\right)^2+m^2\right)\right]}
,
\ee
where $\tilde{\mu}(\lambda)$ is given by
\bea
\mbox{For odd $d+1$}~&:&  \tilde{\mu}(\lambda)
=\frac{1}{2^{d} \pi ^{\frac{d+1}{2}}\Gamma \left(\frac{d+1}{2}\right)}
\prod_{j=0}^{(d-2)/2} (\lambda^2+j^2)
\nonumber\\
\mbox{For even $d+1$}~&:&
\tilde{\mu}(\lambda)=
\frac{1}{2^{d} \pi ^{\frac{d+1}{2}}\Gamma \left(\frac{d+1}{2}\right)}\times \left[
\lambda \tanh{(\pi\lambda)}
\right]\times
\prod_{j=1/2}^{(d-2)/2} (\lambda^2+j^2).
\eea
When $d+1=2$, the product should be omitted.

In order to compare with our previous results, we specialize again to the case $d+1=2$, and use expression (\ref{KtoZ}) to find $\log Z$.  We obtain
\bea\label{eq:H2Zheatkernel}
&&\log Z(\Delta)\nonumber\\
&=&
\frac{-2\mbox{Vol}_{reg}(H_{2})}{(4\pi L^2)}\left[
 \half \mu^2
   -\half \left(  \mu^2+\frac{1}{12 } \right)\log \mu^2
+ \left(\mu^2+\frac{1}{12}
\right)\log(L \Lambda)-\frac{7}{960} \mu^{-2}
\right]+{\cal O}(\mu^{-4})\nonumber\\
\eea
 where we have introduced a lower cutoff of $\epsilon=e^{-\gamma}\Lambda^{-2}$ on the heat kernel integral in Eq.~(\ref{KtoZ})
and ignored a $\mu$-independent but quadratically divergent (in $\Lambda$) terms.
Under regularization of the AdS$_2$ volume, we indeed find that our expression (\ref{Zfinalzero1}) matches this full eigenfunction-expanded heat kernel result
in (\ref{eq:H2Zheatkernel}).  Although we have not shown the calculation here, these two expressions will continue to be equal term by term in powers of $1/\mu$.

\section{Two interesting properties of AdS$_2$ zero modes}
\label{funthings}
Before extending our results to other dimensions or to thermally-identified spacetimes, we pause to study two important properties of the zero modes given by (\ref{AdS2modes}) and (\ref{AdS2zeros}).  First, these modes have a natural physical interpretation as the Wick rotation of quasinormal modes of an AdS$_2$ black hole.  Second, these modes are manifestations of the discrete series representations of $SO(2,1)$ on the space of the $S^1$ parameterized by $\theta$.
For convenience, we set the AdS length to one in this section.
\subsection{Zero modes Wick rotate to quasinormal modes}
\label{Wickrotation}
So far we have been working entirely in Euclidean space; we have really calculated the one-loop determinant for the Euclidean hyperbolic space $H_2$.  If we wish to study the partition function in the context of a Lorentzian spacetime, we can either first study its Euclidean section, or we can locate the poles in the partition function in the complex $\Delta$ plane directly in the Lorentzian spacetime.  As shown in \cite{Denef:2009kn}, the zero-mode method Wick-rotates to a quasinormal mode method, because the zero modes themselves Wick-rotate into quasinormal modes.

We choose coordinates on Euclidean space so $\theta\sim\theta+2\pi$ is the Euclidean time circle and $\rho$ is a radial coordinate such that the time circle shrinks smoothly at the origin $\rho=0$.
The zero mode boundary conditions become
\be
\phi^*_\pm \sim \rho^n e^{\pm i n \theta} \text{ as } \rho \rightarrow 0, \qquad
\phi^*_\pm \sim \rho^{-\Delta} \text{ as } \rho \rightarrow \infty,
\ee
where $n$ is a nonnegative integer.  Under Wick-rotation $\theta \rightarrow 2\pi  T i  t $, the condition at infinity does not change.  Additionally replacing $n=\omega_n/2\pi T$ the condition at $\rho\rightarrow 0$ becomes
\begin{align}
\phi^* &\sim \rho^{\frac{\omega_n}{2\pi T}} \exp\left[\mp\omega_n t\right],
\nn\\
& \sim \exp \left[\omega_n \left(\frac{\log \rho}{2\pi T}\mp t\right)\right],
\nn\\
& \sim \exp \left[-iz^* \left(\frac{\log \rho}{2\pi T}\mp t\right)\right],
\end{align}
where in the last line we set $\omega_n=-iz^*$.  We can now recognize $z^*$ as the (anti)quasinormal mode frequencies, since these modes are purely (out)ingoing here.  For a black hole spacetime these are true (anti)quasinormal modes as the origin of Euclidean space Wick-rotates to the horizon of the corresponding black hole spacetime. Whenever the complex conformal dimension is tuned to $\Delta=\Delta_\star$ so that one of the (anti)quasinormal mode frequencies satisfies
\be
z^* =2 \pi i n T,
\ee
for integer $n$, then the Euclidean section will support a zero mode at $\Delta_\star$, and hence $Z(\Delta)$ will have a pole at $\Delta_\star$.

Specifying to the case of AdS$_2$, the coordinates $\rho=\sinh \eta$ and $\theta$ in (\ref{AdS2EucDisc}) are in fact coordinates of the above type for $T=1/2\pi$. We can thus examine the Wick-rotation of the zero modes in (\ref{AdS2modes}), under $\theta\rightarrow it$.  We also use the similar coordinate $r=\cosh \eta$, related to $\rho$ by $r^2=1+\rho^2$.  In these coordinates the horizon of the Wick-rotated spacetime will be at $r=1$:
\be
ds^2=-\left(r^2-1\right)dt^2+\frac{dr^2}{r^2-1}.
\ee
We recognize this spacetime as an AdS$_2$ black hole.  Here, the modes (\ref{AdS2modes}) become
\be
\phi_{\Delta_\star l} = e^{-l t}\left(1-r^2\right)^{|l|/2} F\left[\Delta_\star+|l|,|l|+1-\Delta_\star; |l|+1; \frac{1-r}{2}\right],
\ee
and these modes satisfy the boundary condition $\phi \sim \rho^{-\Delta_\star}, \, \phi \nsim \rho ^{\Delta_\star-1}$ when
\be
\Delta_\star \in \mathbb{Z}_{\leq 0}, \qquad |l|\leq -\Delta_\star.
\ee
In the Wick rotation above, modes with $l\geq 0$ correspond to quasinormal modes, while those with $l<0$ correspond to antiquasinormal modes.  Hence when identifying with the mode frequency $\omega$ in the Lorentzian spacetime, for $l\geq 0$ we set $l=i\omega$, while for $l<0$ we instead have $l=-i\omega$.

Using this identification and the identity 15.3.30 in \cite{AandS}, we find
\be\label{AdS2qnmodes}
\phi_{\Delta \pm} \propto e^{\mp i\omega t}\rho^{i\omega} F\left[\frac{\Delta+i\omega}{2},\frac{i\omega+1-\Delta}{2};1+i\omega;-\rho^2\right],
\ee
where ($-$)$+$ indicates that we expect a (anti)quasinormal mode.  Near $\rho=0$, the candidate quasinormal modes are purely ingoing, while the antiquasinormal modes are purely outgoing.

While the Euclidean zero modes satisfying the $\phi \sim \rho^{-\Delta_\star}$ boundary condition only exist for values of $\Delta=\Delta_\star$ where $Z(\Delta)$ has a pole, (anti)quasinormal modes are present for any physical $\Delta$. Instead we have to find the particular values of $\omega$ where (\ref{AdS2qnmodes}) satisfies $\phi \sim \rho^{-\Delta}$, and only at special $\Delta_\star$ will $i\omega$ be an integer.

To examine this behavior further, we now fix $\Delta$ be a generic but physical conformal dimension.  Using 15.3.7 in \cite{AandS} to expand (\ref{AdS2qnmodes}) as $\rho\rightarrow \infty$, we find
\begin{align}
\phi_{\Delta\pm} \sim &
e^{\mp i\omega t}\rho^{-\Delta}\frac{\Gamma(1+i\omega)\Gamma\left(\frac{1}{2}-\Delta\right)}
{\Gamma\left(\frac{i\omega+1-\Delta}{2}\right)\Gamma\left(1+\frac{i\omega-\Delta}{2}\right)}
F\left[\frac{\Delta+i\omega}{2},\frac{\Delta-i\omega}{2};\frac{1}{2}+\Delta;\frac{-1}{\rho^2}\right]
\nn\\
& + e^{\mp i\omega t}\rho^{\Delta-1}\frac{\Gamma\left(1+i\omega\right)\Gamma\left(\Delta-\frac{1}{2}\right)}
{\Gamma\left(\frac{\Delta+i\omega}{2}\right)\Gamma\left(\frac{1+i\omega+\Delta}{2}\right)}
F\left[\frac{1-\Delta+i\omega}{2},\frac{1-\Delta-i\omega}{2};\frac{3}{2}-\Delta;\frac{-1}{\rho^2}\right].
\end{align}
For $\Delta$ physical, that is real and positive, $\phi_{\Delta\pm}$ satisfies $\phi \sim \rho^{-\Delta_\star}$ when
\be\label{omegaqn}
\Delta+i\omega \in \mathbb{Z}_{\leq 0}.
\ee
For a given $\Delta$, any $\omega$ satisfying this equation will be a quasinormal frequency. Note that when we are considering physical $\Delta$, our boundary condition at infinity is merely normalizability; when we are working in the complex $\Delta$ plane, we instead insist on removing entirely one of the two behaviors at infinity.

We can now see explicitly that the quasinormal modes in Lorentzian space Wick-rotate into zero modes in Euclidean space, so we can find the poles in $Z(\Delta)$ from either perspective.  To find the poles in a Lorentzian spacetime with a horizon, we first fix a physical $\Delta$. We then find the (anti)quasinormal modes, i.e. modes which are purely (out)ingoing at the horizon and satisfy Dirichlet boundary condition at infinity.  The frequencies of these modes can be written as function $\omega(\Delta)$, which can then be analytically continued to the complex $\Delta$ plane.   Poles in $Z(\Delta)$ are at $\Delta=\Delta_\star$ such that $i\omega(\Delta_\star)\in\mathbb{Z}$, where the integer is positive for quasinormal modes and negative for antiquasinormal modes.

If we instead wish to find the poles in $Z(\Delta)$ directly from Euclidean space, we allow $\Delta$ to range throughout $\mathbb{C}$ from the beginning.  We instead fix a boundary condition of smoothness and singlevaluedness in the interior of the spacetime, and then find the $\Delta_\star$ which satisfy $\phi \sim \rho^{-\Delta_\star}$ as we near the boundary $\rho\rightarrow\infty$.

\subsection{Zero modes as discrete series representations}
Since quasinormal modes are related to the group theory structure of a spacetime, we now consider the group theoretic interpretation of the zero modes (\ref{AdS2modes}).  These modes are manifestations of the discrete series representations of $SO(2,1)$ on the circle $S^1$ parameterized by $\theta$.  Rotating the modes in (\ref{AdS2qnmodes}) back into zero modes in Euclidean AdS$_2$, or $H_2$,
 we obtain
\be
\phi_{\Delta l} = e^{il\theta}(i \sinh \eta)^{|l|}F\left[\frac{\Delta+|l|}{2},\frac{|l|+1-\Delta}{2};1+|l|;-\sinh^2 \eta\right],
\ee
which can also be interpreted as a matrix element of a representation of $SO(2,1)$ with Laplacian eigenvalue set by $-\Delta$.

Before we assign a particular value to $\Delta$, we can see this function is quite similar to a spherical harmonic for $SO(3)$; the $\theta$ dependent piece has weight $l$ under transformations by $\partial_\theta$, the generator corresponding to the $SO(2)$ subgroup. $\Delta$ sets the eigenvalue under the Laplacian, or quadratic Casimir, for the entire $SO(2,1)$.

Next, for the zero modes, $-\Delta_\star$ is restricted to be a positive integer. For the group $SO(2,1)$, representations with $-\Delta_\star \in \mathbb{Z}_{\geq 0}$ are unitary, but with respect to an unusual norm%
\footnote{For further details on special functions on homogenous spaces as representations of associated Lie groups, see \cite{Valenkin2}, especially (12) of 9.4.2 and (4) of 7.4.4.
}. %
These representations are equivalent to the discrete series representations on $SO(2,1)$.  For physical $\Delta$, that is $\Delta$ real and nonnegative, we would not encounter these representations, at least not while considering scalars; fields with spin can have these eigenmodes even for physical $\Delta$, as in \cite{Camporesi:1994ga}.

In our method, where we have continued $\Delta$ into the complex plane, we must consider these zero modes although they are nonphysical.
 Even though they are not square integrable, they do obey our boundary condition $\phi \sim \sinh^{-\Delta}\eta$ at $\eta \rightarrow \infty$, and more importantly they do not have any of the behavior $\sinh^{\Delta-1}\eta$.  Additionally, they do correspond to matrix elements of the discrete series unitary representations on $SO(2,1)$.  In even-dimensional hyperbolic spaces such as $H_2=SO(2,1)/SO(2)$, these discrete representations continue to be present, so we should not be surprised by their appearance at generic $\Delta\in \mathbb{C}$.

\section{The thermal partition function on AdS$_2$}
\label{thermalsection}
We now consider the thermal partition function for Euclidean AdS$_2$ at temperature $T$, where the Euclidean spacetime is constructed from global coordinates
\be\label{globalAdS2}
ds^2=\cosh \sigma ^2 d\tau_E^2+d\sigma^2,
\ee
by identifying the Euclidean time coordinate $\tau_E \sim \tau_E + 1/T$.

Some intuition for the relationship between these coordinates and the Euclidean disc coordinates (\ref{AdS2EucDisc}) can be gained by relating each to the embedding coordinates of the hyperboloid $H_2$ in $\mathbb{R}^{2,1}$.  The embedding measure is $dx_0^2+dx_1^2-du^2$, and the two sets of coordinates are given as
\begin{align}
L \cosh \eta = & \,\,u\, = L \cosh\sigma \cosh\tau_E,
\\
L \sinh \eta \sin \theta = & \,x_0 = L \cosh \sigma \sinh \tau_E,
\\
L \sinh \eta \cos \theta = & \,x_1 = L \sinh\sigma.
\end{align}
In both cases, the Lorentzian spacetime is obtained by Wick rotation of $x_0 \rightarrow i v$.  However, the two rotations are not really equivalent. In the disc coordinates, we rotate $\theta$ which is initially a compact direction.  In the global coordinates, we rotate $\tau_E=i\tau$; global Lorentzian AdS usually refers to allowing $\tau$ to range over all real values.

As the coordinates (\ref{globalAdS2}) make clear, translations in $\tau_E$ are a Killing symmetry.  Thermalizing along this direction is thus quotienting by the subgroup generated by discrete translations of size $T$ in the $\tau_E$ direction.

\subsection{$T$-dependent portion of $Z(\Delta)$ in thermal AdS$_2$}

Working in global coordinates (\ref{globalAdS2}), \cite{Denef:2009kn} finds the temperature dependent piece of the partition function on general thermal AdS spacetimes by first finding the global normal mode frequencies in Lorentzian signature.  The frequencies are ``normal'' because there is no dissipation, so they are purely real:
\be
z_n = \pm \frac{k+\Delta}{L}, \qquad k \in \mathbb{Z}_{\geq 0}.
\ee
These frequencies then correspond to zero modes which manifest as poles in the thermal partition function.  Summing over these poles, \cite{Denef:2009kn} propose
\be \label{ZonlyTparts}
\log Z(\Delta)_{T} = \Pol -2 \sum_{k=0} \log \left(1-e^{-\frac{k+\Delta}{L T}}\right).
\ee
Since the thermal identification is global, not local, the curvature expansion of the heat kernel will not change.  In particular this means that the large $\Delta$ behavior of $Z(\Delta)$ should be temperature independent; we again expect (\ref{likeDHS}), which contains terms of the form $\log \mu$. $Z(\Delta)_T$ in (\ref{ZonlyTparts}), however, does not have any logarithmic terms.  $\Pol$ cannot provide these terms, because it must be a polynomial.

We can resolve this apparent conflict by interpreting $\log Z(\Delta)_T$ as only giving the temperature-dependent behavior of the partition function.  Since the temperature dependence here is a global identification, $T$ will not show up in the local curvature expansion of the heat kernel. If we wish to compute the full partition function on thermal AdS$_2$, and verify that $\text{Pol} (\Delta)$ is in fact a polynomial, then we need to include both contributions.

\subsection{Partition Function on Thermal AdS via Heat Kernel on Quotient of Hyperbolic Space}
From \cite{Gopakumar:2011qs}, using the heat kernel on a quotient of $H_{2}$ constructed from the method of images,
the 1-loop partition function for a scalar field of mass $m$ on thermal AdS$_{2}$ is given by
\bea
\log Z(\Delta)_{T}&=&
-2 \sum_{k=0}
 \log\left[1-e^{-\frac{ k+\Delta}{LT}} \right]
\eea
which agrees with Eq.~(\ref{ZonlyTparts}) without $\Pol$.
Therefore, we see in this concrete example that the normal mode method as used in \cite{Denef:2009kn}
provides only the temperature dependent piece $\log Z(\Delta)_{T}$.

To obtain the full $\log Z(\Delta)$ via a quasinormal or zero mode method, including the local behavior, we need to include the temperature-independent piece. This term is
\bea
&&\int_0^\infty \frac{dt}{t} \int_{H_2/Z} d^2 x \sqrt{g}
K^{H_2}(t,x,x)
=\frac{\mbox{Vol}(H_2/Z)}{\mbox{Vol}_{reg}(H_{2})}\log Z^{H_2}(\Delta)\nonumber\\
&=&\frac{\mbox{Vol}(H_2/Z)}{\mbox{Vol}_{reg}(H_{2})}\left\{
\left[-1+\log \left(L \Lambda _{RG}\right)\right] \Delta (\Delta-1)+\frac{1}{3} \log \left(L\Lambda _{RG}\right)-\frac{1}{4}\right.\nonumber\\
&& \left.\qquad\qquad\qquad
+2\zeta'(-1,\Delta)-(2\Delta-1)\zeta'(0,\Delta)
\right\}.\nonumber\\
\eea
So the {\it full} 1-loop partition function on thermal AdS$_2$ is:
\bea
\log Z^{H_2/Z}(\Delta)&=&
-2 \sum_{k=0}
 \log\left[1-e^{-\frac{ k+\Delta}{LT}} \right]\nonumber\\
 &&+
 \frac{\mbox{Vol}(H_2/Z)}{\mbox{Vol}_{reg}(H_{2})}\left\{
\left[-1+\log \left(L \Lambda _{RG}\right)\right] \Delta (\Delta-1)+\frac{1}{3} \log \left(L\Lambda _{RG}\right)-\frac{1}{4}\right.\nonumber\\
&& \left.\qquad\qquad\qquad
+2\zeta'(-1,\Delta)-(2\Delta-1)\zeta'(0,\Delta)
\right\}.\nonumber\\
\label{H2modZfull}
 \eea
This expression now has the expected logarithmic behavior at large $\Delta$ as seen in the heat kernel curvature expansion, and as expected from the anomaly in even dimensions.  Note that the new term in the second line of (\ref{H2modZfull}) is temperature independent (except for the volume contribution); thus it did not show up in the explicitly temperature-dependent contribution calculated in \cite{Gopakumar:2011qs}, nor in the equivalent calculation done for general AdS in \cite{Denef:2009kn}.

\section{Extension of story to all even dimensions}
\label{2nsection}
We now extend the procedures of Section \ref{d+1=2section} to all even dimensional AdS$_{d+1}$ for $d+1=2n$.

As shown in Appendix \ref{proofofzeromode}, there are modes regular at the origin, with $\sinh^{-\Delta}\eta $ behavior at $\eta \rightarrow \infty$, that solve
\beq\label{Laplgenerald}
\left(-\nabla^2 + \frac{\Delta_\star(\Delta_\star-d)}{L^2}\right)\phi_\star=0.
\eeq
These modes are given by (\ref{phidmodes}), under the restriction
\beq\label{zerocondgenerald}
-\Delta_\star \geq l \geq m_{d-1} \geq \ldots \geq m_2 \geq |m_1|, \qquad {\Delta_\star, l, m_i} \in \ZZZ.
\eeq
This is the same pattern of quantum numbers for spherical harmonics on $S^{d+1}$.  Thus, the degeneracy for the AdS$_{d+1}$ labeled by the integer $k=-\Delta_\star$ is the same as for a spherical harmonic in $S^{d+1}$ with total angular momentum labeled by $k$.

Following \cite{Denef:2009kn}, we denote this degeneracy $Q(k)$ where
\be
Q(k) = \frac{2k+d}{d}\left(\begin{array}{c}
k+d-1\\d-1
\end{array}\right).
\ee
We can then express $Z(\Delta)$ as in (\ref{Zaspoles})
\beq\label{logZsumd}
\log Z(\Delta)  = \Pol -\sum_{k\geq 0} Q(k) \log (k+\Delta),
\eeq
where the conformal dimension is related to the mass via%
\footnote{We only sum over the $+$ solution for $\Delta = d/2 \pm \mu$, because we impose Dirichlet-like boundary conditions at the AdS boundary. Conversely, for the sphere case studied in \cite{Denef:2009kn}, they must sum over both $\Delta = d/2 \pm i\nu$, $\nu \equiv \sqrt{(mL_{dS})^2- (d/2)^2}$.  This difference accounts for the discrepancy between the analytic continuation of sphere results and the actual AdS results, as studied via a Green function method for $d+1=4$ in the appendix of \cite{Camporesi:1991nw}.
}
\beq
\Delta = \frac{d}{2} + \mu, \qquad \mu \equiv \sqrt{m^2L^2 + (d/2)^2}.
\eeq
Defining $\delta_s$ as an operator on a function $f(s)$ via
\be
\delta_s f(s) \equiv f(s-1),
\ee
we regularize (\ref{logZsumd}) as
\be \label{logZzetad}
\log Z_{AdS_{2n}}(\Delta) = \text{Pol}(\Delta) + Q(-\Delta + \delta_s)\zeta'(s,\Delta)|_{s=0}.
\ee
 To finish the computation, we need to compare the large $\mu$ expansion of (\ref{logZzetad}) to the heat kernel curvature expansion as in (\ref{Zfromcurvature}) and (\ref{curvinvar}).  In the next section we will explicitly evaluate $\Pol$ for the specific cases of $d+1=4$, showing agreement with previous computations of $Z(\Delta)$ in both cases.

\subsection{Finding $\Pol$ for $d+1=4$}
In this subsection, we repeat the analysis
in Section \ref{findPolAdS2} for AdS$_4$ to obtain $\Pol$.
From the zero modes obtained in the previous section, we find:
\bea
\log Z(\Delta)
&=& \Pol
+\frac{1}{6} \left(-2 \Delta ^3+9 \Delta ^2-13 \Delta +6\right) \zeta'(0,\Delta)\nonumber
\\
&&
+\left(\Delta ^2-3 \Delta +\frac{13}{6}\right) \zeta'(-1,\Delta)
+\left(-\Delta+\frac{3}{2} \right)  \zeta'(-2,\Delta)
+\frac{1}{3} \zeta'(-3,\Delta),
\eea 
or
\bea \label{logZzetaH4}
\log Z(\mu)  &=&\text{Pol}(\mu) +
\frac{25 }{144}\mu ^4-\frac{1}{16}\mu ^2
+
\frac{1}{5760}\left(-240 \mu ^4+120 \mu ^2+17\right) \log \mu ^2
+\cO(\mu^{-2})~.
\eea
Similarly to what is done for AdS$_2$ in Sec.~\ref{findPolAdS2}, from the heat kernel local curvature expansion on $H_4$, we obtain:
\bea
\log Z(\mu) &=&\frac{\mu^4}{16}
-\frac{1}{48} \left(1+4e^{\gamma } \Lambda ^2  L^2\right)\mu ^2 -\frac{1}{5760}\left(-240 \mu ^4+120 \mu ^2+17\right) \log \left(\Lambda ^2 L^2\right) 
\nonumber\\
&&+\frac{1}{5760}\left(-240 \mu ^4+120 \mu ^2+17\right) \log \mu ^2+\cO(\mu^{-2}).
\eea
Here we have again eliminated a divergent constant term, although we are left with a polynomially divergent term dependent on $\mu^2$.
Matching this expression with the zero-mode expression allows us to find Pol:
\bea
\mbox{Pol}(\mu) &=&
 \half\left[
-\frac{2}{9}+\frac{1}{6}  \log \left(  L \Lambda_{RG}\right)\right]\mu ^4
+\frac{1}{24}\left[
1- \log \left(L \Lambda_{RG} \right)-2e^{\gamma } L^2  \Lambda_{RG} ^2\right]\mu^2\nonumber\\
&&\qquad\qquad-\frac{17}{2880} \log (  L \Lambda_{RG}),\nonumber\\
\Rightarrow \Pol&=&
 \half\left[
-\frac{2}{9}+\frac{1}{6}  \log \left(  L \Lambda_{RG}\right)\right]\left(\Delta-\frac{3}{2}\right) ^4
+\frac{1}{24}\left[
1- \log \left(L \Lambda_{RG} \right)-2e^{\gamma } L^2  \Lambda_{RG} ^2\right]\left(\Delta-\frac{3}{2}\right)^2\nonumber\\
&&\qquad\qquad-\frac{17}{2880} \log (  L \Lambda_{RG}).
\eea
Thus, the full 1-loop partition function on $H_4$ is given by
\bea\label{1loopH4}
&&\log Z(\Delta) \nonumber\\
&=& \half\left[
-\frac{2}{9}+\frac{1}{6}  \log \left(  L \Lambda_{RG}\right)\right]\left(\Delta-\frac{3}{2}\right) ^4
+\frac{1}{24}\left[
1- \log \left(L \Lambda_{RG} \right)-2e^{\gamma } L^2  \Lambda_{RG} ^2\right]\left(\Delta-\frac{3}{2}\right)^2
\nonumber\\
&&-\frac{17}{2880} \log (  L \Lambda_{RG})
+\frac{1}{6} \left(-2 \Delta ^3+9 \Delta ^2-13 \Delta +6\right) \zeta'(0,\Delta)
+\left(\Delta ^2-3 \Delta +\frac{13}{6}\right) \zeta'(-1,\Delta)
\nonumber\\
&&
+\left(-\Delta+\frac{3}{2} \right)  \zeta'(-2,\Delta)
+\frac{1}{3} \zeta'(-3,\Delta).\nonumber\\
\eea

\subsection{Comparison between Thermal partition function in Quasinormal Mode and Heat-Kernel Method}
Proceeding as in Section \ref{thermalsection}, we now compute the 1-loop partition function on thermal AdS$_{2n}$.

According to the ``quasi''normal mode method, using normal modes on Lorentzian global AdS$_4$, the thermal partition function of a complex scalar on
thermal AdS$_4$ is given by
\bea\label{dhs1loopthermalads4}
\log Z(\Delta)_{T}-\Pol
&=&
- 2\sum_{n,l\ge0}(2l+1)\log\left[ 1-e^{-\frac{2n+l+\Delta}{LT}}\right]\nonumber\\
&=&
-  \sum_{k\ge0}(k+1)(k+2)\log\left[ 1-e^{-\frac{k+\Delta}{LT}}\right].
\eea
On the other hand, from \cite{Gopakumar:2011qs}, the 1-loop partition function of a complex massive scalar on thermal AdS$_{d+1}$ using the heat kernel method is
\be\label{heatkernel1loopscalar}
\log Z(\Delta)_{T}= \sum_{p=1}
\frac{2}{e^{-(\frac{d}{2})\frac{p}{LT} }
\left(e^{\frac{p}{ LT}}-1\right)^{d}
}
 \frac{1}{p}
e^{-\frac{p}{LT}\left(\Delta-\frac{d}{2}\right)}.
\ee
In particular, for $d=3$, we obtain
\bea\label{hk1loopthermalads4}
\log Z(\Delta)_{T}&=&\sum_{p\ge 1} \frac{2}{p(1-e^{-\frac{p}{L T}})^3} e^{-\frac{p \Delta}{L T}} \nonumber\\
 &=&
- \sum_{k\ge 0}
(k+1)(k+2)
\log[1-e^{- \frac{(k+\Delta)}{LT}}].
\eea 
So we see that the temperature-dependent piece of the heat-kernel answer in Eq.~(\ref{hk1loopthermalads4}) agrees with the thermal normal mode method answer in Eq.~(\ref{dhs1loopthermalads4}) for AdS$_4$.

Again, as for AdS$_2$, if we wish to show Pol$(\Delta)$ is truly a polynomial, we need to compare the {\it full} partition function for thermal AdS$_4$, i.e. including the temperature-independent part. To do this computation, we add to Eq.~(\ref{hk1loopthermalads4}) the contribution from the zero mode method on $H_4$, which is just Eq.~(\ref{1loopH4}) multiplied by $ \mbox{Vol}(H_4/Z)/{\mbox{Vol}_{reg}(H_{4})}$.

As a side note, in fact we can show that for general AdS$_d$, the quasinormal mode computation (including only thermal normal modes) on thermal AdS$_{d}$ agrees with the heat kernel answer in Eq.~(\ref{heatkernel1loopscalar}):
\bea\label{ZTads4}
\log Z(\Delta)_{T}-\Pol&=&
-2\sum_{p,l\ge0}\frac{2l+d-2}{d-2}{l+d-3 \choose d-3}
\log\left[ 1-e^{-\frac{2p+l+\Delta}{LT}}\right]
\nonumber\\
&=& \sum_{p=1}
\frac{2}{e^{-(\frac{d}{2})\frac{p}{LT} }
\left(e^{\frac{p}{ LT}}-1\right)^{d}
}
 \frac{1}{p}
e^{-\frac{p}{LT} \left(\Delta-\frac{d}{2} \right)}.
\eea
For odd dimensional AdS, the temperature independent term not computed in \cite{Gopakumar:2011qs} is polynomial (see, e.g. Ref.~\cite{Caldarelli:1998wk}).
In the even dimensional case, however, it is not a polynomial and we conjecture that it could be obtained by the zero mode method above (which in the AdS$_4$ case results in Eq.~(\ref{1loopH4})).  If so, by combining the normal modes on thermal AdS with the zero modes we have found here, the quasinormal or zero-mode method can be used to compute the full AdS$_{2n}$ partition function.

\section{Discussions and Conclusions}

We have calculated the scalar one-loop determinant for even dimensional AdS spacetimes via a zero-mode method. In the case of AdS$_2$, these zero modes are the Wick rotation of quasinormal modes for the AdS$_2$ black hole.  In all even dimensions, our results match with previous results using purely heat kernel methods.

Our method is a nontrivial extension of the zero-mode method developed in \cite{Denef:2009kn} to the case of even-dimensional non-compact spaces.  In a compact space such as $S^{2n}$, the condition on the zero modes is normalizability via square integrability.  Since the condition on the complete set of eigenmodes used in the heat kernel eigenfunction expansion is the same, the modes are very similar.  Conversely, in the AdS$_{2n}$ case, the eigenmodes used in the heat kernel computation are constrained to be delta-function normalizable, whereas the zero modes in our zero-mode method are instead constrained by having only $-\Delta$ behavior at infinity.  This application of the boundary condition at infinity for a non-compact space is novel.

For thermal AdS in even dimensions, our results augment those of \cite{Denef:2009kn}, which provided only the temperature-dependent piece of the partition function.  Since the temperature-independent piece of $\log Z(\Delta)$ has logarithmic components, the polynomial $\Pol$ cannot account for these terms.  Instead, in order to compute the partition function for thermal AdS$_{d+1}$ in even dimensions, we should consider both the zero modes of Euclidean $H_{d+1}$ as well as the thermal zero modes arising from compactification in $\tau_E$.  This paper concentrates on the $H_{d+1}$ modes, while \cite{Denef:2009kn} considered the temperature-dependent thermal modes; considering both effects together results in a $Z(\Delta)$ which is meromorphic as expected.

Our interest here is not just in reproducing the AdS even-dimensional calculation, but instead in further developing the zero/quasinormal mode method, with hopes that it may be useful for more general spacetimes.  This particular case does give a few caveats which may be of importance for more general applicability.  
\begin{itemize}
\item The boundary condition of normalizability on the zero modes (\ref{phiconds}) may be delicate.  In the AdS$_2$ case, the zero modes appear for $\Delta<0$, which means the modes satisfying $\phi \sim \sinh^{-\Delta}\eta$ are not normalizable in the usual sense.  
\item A Euclidean spacetime may be Wick-rotatable into Lorentzian signature in more than one way, as we explicitly explored for AdS$_2$ in Section \ref{thermalsection} and discuss for general $d$ in the appendix. To capture all possible zero modes from Lorentzian-signature spacetimes, it may be necessary to consider quasinormal modes visible in each such Wick rotation, as in the thermal AdS case here.  
\item We have highlighted the relationship between zero modes and the group structure of the spacetime.
\end{itemize} 
It turns out in odd dimensions we will need to consider this information.

Thus far we have not discussed odd dimensional AdS.  First, in odd dimensions, the calculation of the thermal partition function follows the (correct) AdS$_3$ case done explicitly in \cite{Denef:2009kn}.  As odd dimensional AdS has no logarithmic terms in the large mass expansion of $\log Z$, the thermal sum as in (\ref{ZonlyTparts}) is sufficient in this case.  If we wish to compute the scalar partition function for $H_{2n+1}$ without a thermal compactification, the entire result is in fact polynomial in $\Delta$, and so can be captured just by setting $\Pol$ according to the heat kernel curvature expansion.

However this means in the odd dimensional case we should not consider the modes in Section \ref{2nsection} to be zero modes.  This is at first a bit surprising, because they appear to satisfy our condition of having only $\sinh^{-\Delta} \eta$ behavior (and no $\sinh^{\Delta-d} \eta$).  As we argue in the appendix, the distinction between the behaviors is not so clear because they differ by an integer power of $\sinh^2 \eta$.  Additionally, the Laplacian equation (\ref{Laplgenerald}) under the restriction (\ref{zerocondgenerald}) has an abelian monodromy group for odd $d+1$, but it is nonabelian for even $d+1$.

Our most important reason to discount these odd-dimensional modes is group theoretic. We expect these modes should not be considered because they are matrix elements for a discrete series representation, but discrete series representations of the motion group for $H_{d+1}$ only exist in even dimensions.

Regardless, the delicacy of deciding what boundary conditions should be applied in order to determine acceptable zero modes when we consider unphysical general complex $\Delta$ is the most important disadvantage of the zero/quasinormal mode method for computing partition functions. Moving into the complex $\Delta$ plane means we are away from the physical $\Delta>0$ case, and so some of our intuition about which modes to include may break down.

Conversely, the major advantage is that we only require information about the location and multiplicity of the zero/quasinormal modes in the complex $\Delta$ plane; we do not need to normalize the modes or provide a complete set of states as in the heat kernel eigenfunction approach of \cite{Camporesi:1994ga}, or compute characters of representations as in \cite{Gopakumar:2011qs}. We believe this advantage is sufficient to motivate further study and application of the method in other spacetimes.

\section*{Acknowledgements}
We would like to thank  F.~Denef, T.~Hartman, S.~Hartnoll, F.~Larsen, P.~Lisb\~{a}o and J.~Maldacena for valuable discussion. C.~K. is supported in part by the US Department of Energy under grant DE-FG02-95ER40899. G.~N. was supported by DOE grant DE-FG02-91ER40654 and the Fundamental Laws Initiative at Harvard.

\appendix
%
\section{Finding zero modes in AdS$_{d+1}$ for $d+1$ even}\label{proofofzeromode}
We work in spherical coordinates
\beq
ds_E^2 = d\eta^2 + \sinh ^2\eta d\Omega^2_{d},
\eeq
where $d\Omega^2_d$ is the line element of $S^d$.
We are looking for eigenfunctions of the Laplacian satisfying
\be\label{phiconds}
\left(-\nabla^2+\frac{\Delta(\Delta-d)}{L^2}\right)\phi=0, \qquad \phi \sim_{\eta \rightarrow \infty} \sinh^{-\Delta} \eta.
\ee
Following \cite{Camporesi:1994ga}, the eigenfunctions of the Laplacian regular at the origin are given by
\begin{align}\label{phiregular}
\phi &=Y_{l,\vec{m}} q_{l}
\\\nn
q_{l\lambda} &=
 (i \sinh \eta)^l F\left[i\lambda+\frac{d}{2}+l,-i\lambda+\frac{d}{2}+l;l+\frac{d+1}{2};-\sinh^2\left(\frac{\eta}{2}\right)\right].
\end{align}
where $Y_{l,\vec{m}}$ is a spherical harmonic on $S^d$ whose total angular momentum is set by the nonnegative integer $l$. $\vec{m}$ is a set of $d-1$ integers with $l\geq m_{d-1} \geq \ldots \geq |m_1|$, $m_1 \in \mathbb{Z}$. For the specific case of $d+1=2$ dimensions, there is no $\vec{m}$; instead for every nonnegative integer $l$ we have two modes:  $\exp{(il\theta)}$ and $\exp{(-il\theta)}$.

We want to rewrite the eigenvalue as $\Delta(\Delta-d)/L^2$.  Reparameterizing the hypergeometric function in terms of $\Delta$, we find
\beq\label{phidmodes}
\phi_{l\Delta} \propto Y_{l,\vec{m}} (i \sinh \eta)^l F\left[l+\Delta,l+d-\Delta; l+\frac{d+1}{2}; \frac{1-\cosh \eta}{2}\right].
\eeq
Using the quadratic transformation 15.3.30 in \cite{AandS}, we rewrite $q$  (now dropping the $Y$) as
\be
q_{l\Delta} \propto (i \sinh \eta)^l F\left[\frac{l+\Delta}{2},\frac{l+d-\Delta}{2};l+\frac{d+1}{2};-\sinh^2 \eta\right].
\ee
As long as $\Delta -d/2$ is not an integer, we can use 15.3.7 in \cite{AandS} to expand $q$ near $\eta \rightarrow \infty$:
\begin{align}\label{phiatinf}
q_{l\Delta} \sim & (\sinh\eta)^{-\Delta} \frac{\Gamma\left(l+\frac{d+1}{2}\right)\Gamma\left(\frac{d}{2}-\Delta\right)}
{\Gamma\left(\frac{l+d-\Delta}{2}\right)\Gamma\left(\frac{l+d+1-\Delta}{2}\right)}
F\left[\frac{l+\Delta}{2} , \frac{1+\Delta-d-l}{2} ; 1+\Delta-\frac{d}{2} ; -\sinh ^{-2} \eta\right]
\\\nn
&+ (\sinh\eta)^{\Delta-d}
\frac{\Gamma\left(l+\frac{d+1}{2}\right)\Gamma\left(\Delta-\frac{d}{2}\right)}
{\Gamma\left(\frac{l+\Delta}{2}\right)\Gamma\left(\frac{l+1+\Delta}{2}\right)}
F\left[\frac{l+d-\Delta}{2} , \frac{1-l-\Delta}{2} ; 1-\Delta+\frac{d}{2} ; -\sinh ^{-2} \eta\right].
\end{align}
These modes already satisfy the condition of regularity in the interior of $H_{d+1}$; this is how we chose to consider the solution (\ref{phiregular}) to the second-order differential equation in (\ref{phiconds}).  In order to satisfy the $\eta \rightarrow\infty$ boundary condition in (\ref{phiconds}), we must choose $\Delta$ such that the second line in (\ref{phiatinf}) vanishes. This occurs whenever one of the $\Gamma$ functions in the denominator has a pole; that is, whenever $l+\Delta \in \mathbb{Z}_{\leq 0}$.  Since $l$ is already restricted to be a nonnegative integer as it labels the total angular momentum of the spherical harmonic $Y_{l\vec{m}}$ in (\ref{phiregular}), this forces $\Delta$ to be a negative integer.

For $d+1$ even, $\Delta-d/2$ is not an integer, and our analysis is complete; there is a zero mode for $\Delta_\star \in \mathbb{Z}_{\leq 0}$ with degeneracy equal to the sphere degeneracy for a mode on $S^{d+1}$ with total angular momentum $-\Delta_\star$.

Specifying to the case $d+1=2$, we find
\be\label{d+1=2modes}
\phi_{\Delta l}=  e^{il\theta}(i \sinh \eta)^{|l|} F\left[\Delta+|l|, |l|+1-\Delta;|l|+1;-\sinh^2 \left(\frac{\eta}{2}\right)\right],
\ee
has only $(\sinh \eta)^{-\Delta}$ (and no $(\sinh \eta)^{1-\Delta}$) behavior at $\eta \rightarrow \infty$ whenever $\Delta=\Delta_\star$ for $\Delta_\star+|l| \in \ZZZ$ and $\Delta_\star+|l| \leq 0$.  Since here we are allowing $l \in \ZZZ$ to account for both $\exp (il\theta)$ and $\exp (-il\theta)$,
we can rewrite this requirement as
\be\label{d+1=2cond}
\Delta \in \ZZZ_{\leq 0}, \qquad l \in \ZZZ, \qquad |l| \leq -\Delta.
\ee

The case of odd dimensions is more complicated. $d$ is an even integer, and the analysis using (\ref{phiatinf}) is insufficient because the formula 15.3.7 in \cite{AandS} is invalid when $\Delta-d/2 \in \ZZZ$.  In this case, the top line of (\ref{phiatinf}) differs from the bottom by an integer power of the argument of $F$, $-\sinh^2 \eta$, so we cannot truly separate the two series from each other. The expansion of the top line can contaminate that of the bottom line.

If we do continue by studying the degenerate hypergeometric equation (see \cite{Bateman},\cite{Vidunas2007}), we will in fact find that $\phi_{l\Delta_\star}$ as in (\ref{phidmodes}) with $\Delta_\star+l\in \ZZZ_{\leq 0}$ become polynomials in $\sinh^2\eta$ with only positive powers. As such, $\phi_{l\Delta_\star}$ will contain ``none'' of the behavior $\sinh^{\Delta-d} \eta$, which would have to be a negative power for $\Delta \leq 0$.  As we argue in the conclusion, we nonetheless do not need to include these putative zero modes in the sum for odd dimensions.

As further evidence for the difference between even and odd modes, we again mention that the modes in (\ref{phidmodes}) are matrix elements of the discrete series representation for $SO(d+1,1)$.  These representations should only be unitary in the case of $d+1$ even; again for more details see \cite{Valenkin2}.

\section{Euclidean disc vs. Global coordinates}

We begin with the embedding coordinates for $H_{d+1}$:
\be
\sum_{i=1}^{d}x_i^2 +x_0^2-u^2=-L^2,
\ee
with line element
\be
ds^2 = \sum_{i=1}^{d} dx_i^2+dx_0^2 -du^2.
\ee
To recover the Euclidean disc, we choose
\begin{align}
u & = L \cosh \eta
\\
x_0 & = L\sinh\eta \sin\theta_0
\nn\\
x_1 & = L\sinh\eta \cos\theta_0 \cos \theta_1
\nn\\
x_2 & = L\sinh\eta \cos\theta_0 \sin\theta_1 \cos \theta_2
\nn\\
\dots
\nn\\
x_{d-1} & = L\sinh\eta \cos\theta_0 \left(\prod_{i=1}^{d-2} \sin\theta_i \right) \cos\theta_{d-1}
\nn\\
x_d & = L\sinh\eta \cos\theta_0 \prod_{i=1}^{d-1} \sin\theta_i.
\nn
\end{align}
To recover the usual measure on $d\Omega_d^2$, we should additionally take $\theta_0=\pi/2-\theta$. Wick rotating $\theta_0\rightarrow it$ produces the black hole coordinates for the case of AdS$_2$.  This rotation also sets $x_0 \rightarrow i v$ in the embedding coordinates.

For $d>1$, there is a bispherical set of coordinates on $S^d$ which are useful for Wick rotation:
\begin{align}
u & = L \cosh \eta
\\
x_0 & = L \sinh\eta \sin\phi_1 \sin\phi_0
\nn\\
x_1 & = L \sinh\eta \sin\phi_1 \cos\phi_0
\nn\\
x_2 & = L \sinh\eta \cos\phi_1 \cos\theta_2
\nn \\
x_3 & = L \sinh\eta \cos\phi_1 \sin\theta_2 \cos\theta_3
\nn \\
\ldots
\nn\\
x_{d-1} & = L \sinh\eta \cos\phi_1 \left(\prod_{i=2}^{d-2}\sin\theta_i\right) \cos\theta_{d-1}
\nn\\
x_d & = L \sinh\eta \cos\phi_1 \prod_{i=2}^{d-1} \sin\theta_i.
\end{align}
The metric for these coordinates is
\be
ds^2_{d+1} = L^2 \left(d\eta^2 +\sinh^2 \eta d\phi_1^2 + \sinh^2\eta \sin^2 \phi_1 d\phi_0^2 +\sinh^2 \eta \cos^2\phi_1 d\Omega^2_{d-2}\right).
\ee
From these coordinates, we can again Wick rotate, now setting $\phi_0 \rightarrow i t$ and defining $r_{dS}\equiv \cos{\phi}_1$:
\be
ds^2_{d+1}=L^2d\eta^2 +L^2\sinh^2\eta\left[  \frac{dr_{dS}^2}{1-r_{dS}^2} -  (1-r_{dS}^2) dt^2 + r_{dS}^2 d\Omega^2_{d-2}\right],
\ee which is just a foliation of AdS$_{d+1}$ with the $d$-dimensional static de Sitter space.
 It is very curious that the modes in (\ref{phidmodes}) Wick rotate to ``quasinormal modes'' of this foliation of AdS$_{d+1}$. It would be interesting to explore further the interpretation of these ``quasinormal modes''.

In fact, for an even-dimensional AdS$_{2n}$ space, we can pair the embedding coordinates $x_0$ through $x_{2n-1}$. It is possible to pick coordinates so that we can Wick-rotate easily by the angle between each of the $n$ pairs of coordinates.  In fact written in these bi$^n$ spherical coordinates, we explicitly have written a full commuting set of generators for AdS$_{2n}$.

For odd AdS$_{2n+1}$, such a pairing isn't possible. To get a maximal set of commuting generators we must include one noncompact generator (such as $\tau_E$ below).  It is this difference between even and odd AdS that causes even AdS to have discrete series representations, while odd AdS does not.

For the global coordinates in any dimension, we choose
\begin{align}
u & = L \cosh \sigma \cosh \tau_E
\\
x_0 & = L\cosh \sigma \sinh \tau_E
\nn\\
x_1 & = L\sinh\sigma \cos \theta_1
\nn\\
x_2 & = L\sinh\sigma  \sin\theta_1 \cos \theta_2
\nn\\
\dots
\nn\\
x_{d-1} & = L\sinh\sigma  \prod_{i=1}^{d-2} \sin\theta_i \cos\theta_{d-1}
\nn\\
x_d & = L\sinh\sigma  \prod_{i=1}^{d-1} \sin\theta_i.
\nn
\end{align}
In this case, we Wick rotate by taking $\tau_E \rightarrow i \tau$, which takes $x_0 \rightarrow i v$ as before.  This rotation produces the usual global coordinates on Lorentzian AdS$_{d+1}$ for general $d$.  We can see that $\partial_{\tau_E}$ is a noncompact generator, so this Wick rotation is fundamentally different from those considered above.

Thermal AdS usually refers to global AdS with time identification; that is, setting $\tau_E \sim \tau_E+ 1/T$ for some temperature $T$.  In Lorentzian signature this can more properly be thought of as taking $\tau$ to the covering space (that is, allowing $-\infty < \tau< \infty$) and then identifying up to $1/T$.
\bibliography{1loopandQNMarxivV2}
\end{document}